\documentclass[11pt]{article}
\usepackage{blois2002,epsfig}
\def\be{\begin{equation}}
\def\ee{\end{equation}}
\def\bea{\begin{eqnarray}}
\def\eea{\end{eqnarray}}

\begin{document}
\vspace*{4cm}
\title{C, P, T ARE BROKEN. WHY NOT CPT?}

\author{ L.B. OKUN }

\address{ITEP, Moscow, 117218, Russia}

\maketitle\abstracts{Classification of effects of violation of
seven symmetries -- C, P, T, CP, PT, TC, and CPT -- is discussed.
A graphic mnemonic scheme -- CPT-cube -- is suggested and
illustrated by simple examples. }

\section{Introduction}

C, P, and T symmetries are known to hold at very high accuracy for
electromagnetic and strong interactions \cite{1}. For weak
interactions the 100\% breaking of P and C was discovered in 1957
\cite{2,3,4,5,6,7,8,9}  and served as a cornerstone of electroweak
theory \cite{10}.

The CP violation was discovered as a tiny (milliweak) effect in
the decays of neutral kaons in 1964 \cite{11} and recently in the
decays of $B$ mesons \cite{12}. But we still do not know the
origin(s) of the CP violation.

T violation was explicitly proved for neutral kaons \cite{13,13a} in
accord with CPT invariance \cite{14,15} which remains unshaken in
spite of improved precision of experimental tests and increased
number of theoretical speculations.

As for gravity, it is difficult to suggest an experimental test of
C, P, or T in a classical weak gravitational field
\cite{16,17,18}, while effects of quantum gravity belong to the
big bang which sets the arrow of time in the universe \cite{19},
or to the black holes.

In this talk I will not speak about gravity and will try to
classify effects of violation of seven symmetries -- C, P, T, CP,
PT, TC, and CPT -- in laboratory experiments.

The subject has a vast literature. I apologize for making no
attempt to cover it. My aim is to suggest a simple mnemonic
approach.

\section{CPT-cube}

Let us start by considering three orthogonal axes representing
violation of C, P and T (Fig. 1).

Point 0 at the origin of coordinates corresponds to the
interactions, which are C even, P even, T even, and hence CP even,
PT even, TC even, and CPT even.

Point 1 corresponds to the interactions which are C odd, P even, T
even, and hence CP odd, PT even, TC odd, and CPT odd.

Point 2: C even, P odd, T even, CP odd, PT odd, TC even, and CPT
odd.

Point 3: C even, P even, T odd, CP even, PT odd, TC odd, and CPT
odd.

As a result we have the first four vertices of the CPT-cube. Let
us now consider the other four vertices (Fig. 2): three in the
planes CP (point 4), PT (point 5), TC (point 6), and the last one
-- outside these planes (point 7).

Point 4: C odd, P odd, T even, CP even, PT odd, TC odd, and CPT
even.

Point 5: C even, P odd, T odd, CP odd, PT even, TC odd, and CPT
even.

Point 6: C odd, P even, T odd, CP odd, PT odd, TC even, and CPT
even.

Point 7: C odd, P odd, T odd, CP even, PT even, TC even, and CPT
odd.

Thus for each of the seven transformations there are four even
vertices and four odd.

For C: 0, 2, 5, 3 are even; 1, 4, 7, 6 are odd (the rear and front
sides of the cube, correspondingly).

For P: 0, 3, 6, 1 are even; 2, 5, 7, 4 are odd (the left and the
right sides, correspondingly).

For T: 0, 1, 4, 2 are even; 3, 6, 7, 5 are odd (the lower and
upper sides< correspondingly).

For CP: 0, 4, 7, 3 are even; 1, 2, 5, 6 are odd (two vertical
diagonal planes).

For PT: 0, 5, 7, 1 are even; 2, 3, 6, 4 are odd (two diagonal
planes orthogonal to the page).

For TC: 0, 2, 7, 6 are even; 1, 3, 5, 4 are odd (two diagonal
planes whose intersection is parallel to the page).

For CPT: 0, 4, 5, 6 are even; 1, 2, 3, 7 are odd (four diagonals
of lower and upper planes).

The above classification can be easily memorized 
by assigning to each point the corresponding
signature of (C, P, T) with + for even and - for odd:
0(+,+,+), 1(--,+,+), 2(+,--,+), 3(+,+,--), 4(--,--,+),
5(+,--,--), 6(--,+,--), 7(--,--,--).

\section{C and P violation with CP conservation}

Let us start with the interaction of $V-A$ weak charged currents
discovered in 1957. The products $VV$ and $AA$ belong to point 0,
while the product $VA$ -- to point 4,
as the vector current $V$ is C, P, and T odd, while
the axial current $A$ is C even, P even, and T odd.
The experimental
manifestation of the latter is seen directly in the decay $K_1 \to
2\pi$, while of the former in the decay $K_2 \to 3\pi$, where
$K_1$ and $K_2$ are correspondingly C odd and C even
superpositions of $K^0$ and $\bar K^0$:
\begin{equation}
K_1 = \frac{1}{\sqrt{2}}(K-\bar K) \; , \;\; K_2 =
\frac{1}{\sqrt{2}}(K+\bar K) \;\; . \label{1}
\end{equation}
As both $K_1$ and $K_2$ are P odd (pseudoscalars), $K_1$ is CP
even, while $K_2$ is CP odd.

The $2\pi$ in $J=0$ state is C even, P even, and CP even. Hence
the decay $K_1^0 \to 2\pi$ is C odd, P odd, but CP even. As for
the state of $3\pi$ with $J=0$, it is P odd independent of values
of relative angular momenta $l$ and $L$. The dominant state with
$L =l=0$ is C even. Therefore the corresponding decay $K_2^0 \to
3\pi$ is mainly C even, P even and CP even.

The manifestation of $VA$ (point 4) in the spin-momentum angular
correlations in $\beta$-decay, decays of $\mu$ and $\tau$ leptons,
semileptonic decays of mesons and nonleptonic decays of hyperons
is impossible without interference with terms $VV$ and $AA$ (point
0) in the square of modulus of amplitude.

The same refers to the P violating and C violating correlations
induced by neutral currents which were discovered in the 1970's.

All the above processes are mediated by virtual $W$ and $Z$
bosons. After the discovery of these bosons in the early 1980's a
lot of experimental data has been collected on C and P violating
asymmetries in their production and decay processes caused by
interference of points 4 and 0.

\section{CP and T violation}

The discovery in 1964 of the $2\pi$ decays of the long-lived
neutral kaons revealed that CP is violated. For almost three
decades the effective interaction responsible for these decays was
consistent with point 6: C odd, P even, T odd transition in vacuum
of $K_2$ into $K_1$, described by a complex parameter $\varepsilon$:
\begin{equation}
K_S = K_1 + \varepsilon K_2 \; , \;\; K_L = K_2 + \varepsilon K_1
\;\; . \label{2}
\end{equation}

The presence of $\varepsilon$ in eqs.(\ref{2}) leads to the
conclusion that the probability of transformation of $K$ into
$\bar K$ during a time interval $t$ is not equal to the
probability of transformation of $\bar K$ into $K$ during the same
time interval. (The two amplitudes are different
because of the presence of two different time exponents,
describing the propagation of $K_L$ and $K_S$.) 
This prediction of violation of time reversal was
experimentally  confirmed \cite{13} only a few years ago.
Another confirmation of the time reversal violation
came from a CP and T violating asymmetry between 
$e^{+}e^{-}$ planes and $\pi^{+}\pi^{-}$ planes observed
\cite{13a} in the decays $K_L \to \pi^{+}\pi^{-}e^{+}e^{-}$.

Only recently a consensus has been reached on the value of another
parameter, $\varepsilon^{\prime}$, describing the direct decay of
$K_2$ into $2\pi$ (point 5, which is C even, P odd and T
odd).

Point 5 is also responsible  for the dipole electric moments of
such particles as neutron and electron the search for which has up to
now brought no positive evidence. (The term
$\mbox{\boldmath$\sigma$}{\bf E}$, where
$\mbox{\boldmath$\sigma$}$ represents the spin of the particle,
while ${\bf E}$ electric field, is C even, P odd, T odd; 
so that the interaction is CPT invariant.)

Let us stress that at the level of the standard electroweak gauge
lagrangian both points 5 and 6 for hadrons stem from the same
origin, namely the phase of the CKM-matrix of charged currents of
quarks.

Of special interest in connection with point 5 is the vanishingly
small upper limit on the so-called $\theta$ term in QCD:
\begin{equation}
{\cal L}_{\theta} = \theta G_{\mu\nu} G_{\rho\sigma}
\varepsilon^{\mu\nu\rho\sigma} \;\; , \label{3}
\end{equation}
where $G_{\mu\nu}$ is the gluonic field tensor.

\section{CP violating charge asymmetries}

Both points 5 and 6 must manifest themselves
in charge asymmetries. Thus from
eq. (\ref{2}) it follows immediately that the widths of the
semileptonic decays $K_L \to e^+ \nu_e \pi^-$ and $K_L \to e^-
\bar\nu_e \pi^+$ must be different, the effect being proportional
to $2Re\varepsilon$ (point 6). Such charge asymmetry was measured
for both electronic and muonic channels.

Another charge asymmetry had been predicted by Okubo \cite{20}
even before CP-violation was discovered, but is still beyond the
reach of experiments. According to Okubo,
\begin{equation}
\frac{\Gamma(\Sigma^+ \to p \pi^0)}{\Gamma(\Sigma^+ \to n \pi^+)}
\neq \frac{\Gamma(\bar\Sigma^- \to \bar p
\pi^0)}{\Gamma(\bar\Sigma^- \to \bar n \pi^-)} \;\; . \label{4}
\end{equation}
In order to show how this effect appears let me recall that for
amplitudes of both S and P waves the following equalities hold
\begin{eqnarray}
A(\Sigma^+ \to p\pi^0) & = & \sqrt{\frac{2}{3}}A_3 -
\sqrt{\frac{1}{3}}A_1 \;\; , \nonumber \\ A(\Sigma^+ \to n\pi^+) &
= & \sqrt{\frac{1}{3}}A_3 + \sqrt{\frac{2}{3}}A_1 \;\; , \label{5}
\end{eqnarray}
where $A_3$ and $A_1$ are amplitudes for final states with isospin
$T=3/2$ and $T=1/2$ correspondingly. Similar relations hold for
antiparticles:
\begin{eqnarray}
\bar A(\bar\Sigma^- \to \bar p\pi^0) & = & \sqrt{\frac{2}{3}}\bar
A_3 - \sqrt{\frac{1}{3}}\bar A_1 \;\; , \nonumber
\\
\bar A(\bar\Sigma^- \to \bar n\pi^-) & = & \sqrt{\frac{1}{3}}\bar
A_3 + \sqrt{\frac{2}{3}}\bar A_1 \;\; , \label{6}
\end{eqnarray}
For simplicity let us consider only S wave amplitudes. In doing so
we do not lose generality because S and P waves do not interfere
in the expressions for partial widths.

The moduli of isotopic amplitudes, as well as the final state
interaction phase shifts, are the same for particle and
antiparticle ($\bar A_1 = A_1$, $\bar A_3 = A_3$, $\bar\delta_1 =
\delta_1$, $\bar\delta_3 = \delta_3$), while the CP violating
phases  have opposite signs ($\bar\Delta = -\Delta_1$,
$\bar\Delta_3 - =\Delta_3$). As a result we get inequality
(\ref{4}) if $\delta_3 \neq \delta_1$ and $\Delta_3 \neq
\Delta_1$. (The former condition is valid in the standard
electroweak theory \cite{200,201}, while the latter is known to
hold from the $\pi N$-scattering experiments.)

A similar reasoning was applied by Sakharov \cite{21}  when in
1967 he addressed the problem of baryonic asymmetry of the
universe. He assumed CP violation for baryon number violating
processes in order to get different cross-sections for specific
processes with nucleons and antinucleons. The difficulties in
working out a consistent theory of baryogenesis have directed
theoretical thinking towards leptogenesis caused by CP-violation
in leptonic sector, including neutrino oscillations.

\section{Testing CPT with antiparticles}

The faith in CPT invariance is based on quantum field theory, in
particular on locality  of the lagrangian, its Lorentz invariance
and hermiticity.  QFT might be an effective approximate
manifestation of a more fundamental (superstring?) theory. But,
first, I am unaware of any rigorous proof that superstrings
violate CPT and, second, QFT remains the only solid basis of our
phenomenology. Independently of the possible origin of CPT
violation, in order to confront the speculations and experiment
one has to formulate the phenomenological predictions by using the
language of the quantum field theory.

Most of the phenomena suggested for testing CPT belong to the
point 1: they are C-odd, but PT-even. Examples are:
\begin{enumerate}
\item the search for mass differences of particles and
corresponding antiparticles: $m_{K^0} -m_{\bar K^0}$, $m_{K^+} -
m_{K^-}$, $m_{e^-} -m_{e^+}$, $m_{\mu^-} -m_{\mu^+}$, $m_n
-m_{\bar n}$, $m_{\nu_e} -m_{\bar\nu_e}$, etc.
\item the search for nonvanishing sum of magnetic moments of a
particle and its antiparticle: $\mu_{\mu^+} \neq -\mu_{\mu^-}$,
$\mu_{e^+} \neq -\mu_{e^-}$, $\mu_p \neq -\mu_{\bar p}$, etc.
\end{enumerate}

Especially popular nowadays are speculations on non-vanishing mass
differences between neutrinos and
antineutrinos\cite{22,23,24,25,26,27,28,29,32a,30,31}. Most of
them (though not all) are linked with the violation of Lorentz
invariance and/or locality. The most recent references are
triggered by the LSND anomaly \cite{32}. Beware! The CPT violation
in neutrino sector induces at the level of radiative corrections
observable effects among charged leptons, where high precision
tests of the CPT symmetry are available \cite{33}.

A CPT violating effect due to interference of points 3 and 0 would
be muon polarization 
perpendicular to the decay planes of $K_L^0
\to \mu^+ \nu_{\mu}\pi^-$ and $K_L^0 \to \mu^-
\bar\nu_{\mu}\pi^+$, if it turns out to be the same for 
both decays. In that case,
the correlation ${\bf s}_{\mu}[{\bf k}_{\mu}
\times {\bf k}_{\pi}]$ is C even, P even, but T odd. Hence it is
CPT odd. However ``a fake T violation'' could be caused by the
final state muon-pion scattering \cite{34,35} at point 0 with
phase $\delta \sim \alpha/3$. The experimental upper limit for
such polarization is 0.5\% \cite{36}.

Note that the same transverse polarization in the decays $K^+ \to
\mu^+ \nu_\mu \pi^0$ and $K^- \to \mu^- \bar\nu_\mu \pi^0$ cannot
be faked by a simple final state electromagnetic scattering.

As an example of manifestation of point 7 let us consider the
electric dipole moments of a particle and its antiparticle, say,
$e^-$ and $e^+$, or $\nu_e$ and $\bar\nu_e$.

If P and T are broken, while C is conserved (point 6), the
electric dipole moments are nonvanishing, however their sum must
vanish, because they (similarly to charges and to the ordinary
magnetic dipole moments described by point 0) must have opposite
signs. This can be easily seen from the negative C parity of the
photon.

If we now consider point 7, we see that it must be not only P and
T odd but also C odd. This requires a term  which gives not only
the same absolute value, but also the same sign to the electric
dipole moments of a particle and its antiparticle. If both
points 7 and 6 are present, then $d_{\nu_e}\neq d_{\bar\nu_e}$.
Similar inequalities would be valid for all leptons and quarks and
hence hadrons.

\section{CPT and hermiticity}

What would the discovery of circular polarization of a photon
from the decay $\pi^0 \to \gamma\gamma$ (or $\eta^0 \to \gamma\gamma$)
mean? 
One can easily see that
this would be a signal that CPT is broken. Indeed, C is conserved
in this decay, while the product ${\rm\bf s k}$ is P odd, but T
even. (Here ${\rm\bf s}$ and ${\rm\bf k}$ are photon's spin and
momentum respectively; ${\rm\bf s}$ is P even, ${\rm\bf k}$ is
P odd, while both are T odd. Similar considerations apply
to the longitudinal polarizations of muons in the decay
$\eta^{0} \to \mu^{+}\mu^{-}$ observed at the
level of 6$\times$10$^{-6}$. The correlation 
${\rm\bf s_{\mu} k_{\mu}}$ is P odd and T even. On the other hand
it must be C even, because $\eta^{0}$ is C even as well as 
the pair of muons both in the scalar and pseudoscalar states.
Thus observation of the ${\rm\bf s_{\mu} k_{\mu}}$ correlation
would signal the CPT violation.)

Let us consider the two terms in the lagrangian the interference
of which can lead to the correlation ${\rm\bf s k}$
in the decay $\pi^0 \to \gamma\gamma$. First of them
is a scalar $g \phi
F_{\mu\nu}F_{\rho\sigma}\varepsilon^{\mu\nu\rho\sigma}$, the
second one is a pseudoscalar, $h \phi F_{\mu\nu} F^{\mu\nu}$,
where $\phi$ is the pseudoscalar field of the pion, which is C
even, P odd and T odd.

Due to hermiticity of the lagrangian both $g$ and $h$ must be
real. However the degree of circular polarization is proportional
to $g^* h - gh^*$. Hence for a hermitian lagrangian it should
vanish at the tree approximation. (It would appear due to the
difference of the absorptive parts of Fig. 3 in scalar and
pseudoscalar amplitudes.) If the degree of polarization turns out
to be different from that predicted by Fig. 3, e.g. much larger,
we have to break hermiticity.

Let us consider first the case when ${\rm Im} g = 0$, ${\rm Re} h
=0$. In that case the term proportional to $g$ belongs to point 0,
while the term proportional to $h$ belongs to point 2 because it
is C even, P odd, T even, and hence CPT odd.

In the case when ${\rm Re} g =0$, ${\rm Im} h =0$ the $g$-term is
C even, P even, T odd and hence CPT odd (point 3), while the
$h$-term is C even, P odd, T odd, and hence CPT even (point 5).
Interference of points 3 and 5 (as well as that of points 2 and 0)
gives P-violating and CPT-violating correlation ${\rm\bf s k}$.

In the general case, when both $g$ and $h$ are complex, $$ g^* h -
gh^* = 2i {\rm Re}g \cdot {\rm Im}h - 2i {\rm Im}g \cdot {\rm Re}h
\;\; . $$ Here the first term on the r.h.s. represents
interference of points 2 and 0, while the second - of points 3 and
5.

As was noticed by M.B. Voloshin (private communication), the
graphs of Fig. 4 with complex $g$ give complex magnetic moment of
the proton. Similar graphs with complex $h$ give complex electric
dipole moment. Interaction of these moments with corresponding
external fields ${\rm\bf B}$ and ${\rm\bf E}$ would give complex
energy. Complex energy looks rather exotic, but it is not more
exotic than the violation of unitarity of $S$-matrix, or in other
words, the non-conservation of probability. (Note that a complex
energy might correspond to a tunneling of particle through a
wormhole to another vacuum.)

To my knowledge, nobody has tried to measure the helicity of
photons from $\pi^0 \to\gamma\gamma$. 
V.L.~Telegdi (private
communication) noticed that the search for the circular
polarization of photons from the decay of parapositronium is much
easier than from the decay of pions. (At low energy the Compton
scattering is dominant, hence the magnetized iron is a more
effective detector of helicity of low energy photons.)

When thinking about possible mechanism of CPT-violation in the two
photon decay of positronium one sees that circular polarization of
photons might be produced by imaginary electric dipole moment of
electron (see Figs. 5b and 5c) via interference with the standard
QED diagram (Fig. 5a).

Unlike the case of particle-antiparticle mass differences the
above examples of CPT-violating polarizations can be formulated in
a Lorentz-invariant way. In the lagrangian language the
CPT-violating terms have the wrong phase which violates
hermiticity. The antihermitian terms in lagrangian  break
unitarity of $S$-matrix. But testing unitarity directly might be
more difficult than searching for CPT violating correlations
discussed above.

\section{Conclusions}

It is time now to answer the question in the title of this talk:
Why not CPT?

Of course, whatever could be measured should be measured and
whatever could be tested should be tested. There should be no
reservations: such a fundamental symmetry as CPT should be tested.

However one should keep in mind that unlike breaking of C, P, T,
CP, PT, and TC, the breaking of CPT is non-compatible with the
standard quantum field theory, the only basis for a
self-consistent phenomenological description of any process, which
we know up to now. Therefore the chances that CPT breaking would
be discovered are vanishingly small.

I am grateful to S. Bulanov, T. Nakada, B. Roe, V.~Telegdi,
L.~Wolfenstein, and especially M.~Voloshin  for helpful
discussions. This work has been partly supported by RFBR grant No.
00-15-96562, by CERN TH Division, and by A. von Humboldt award.

\newpage

\begin{center}
\includegraphics[width=60mm]{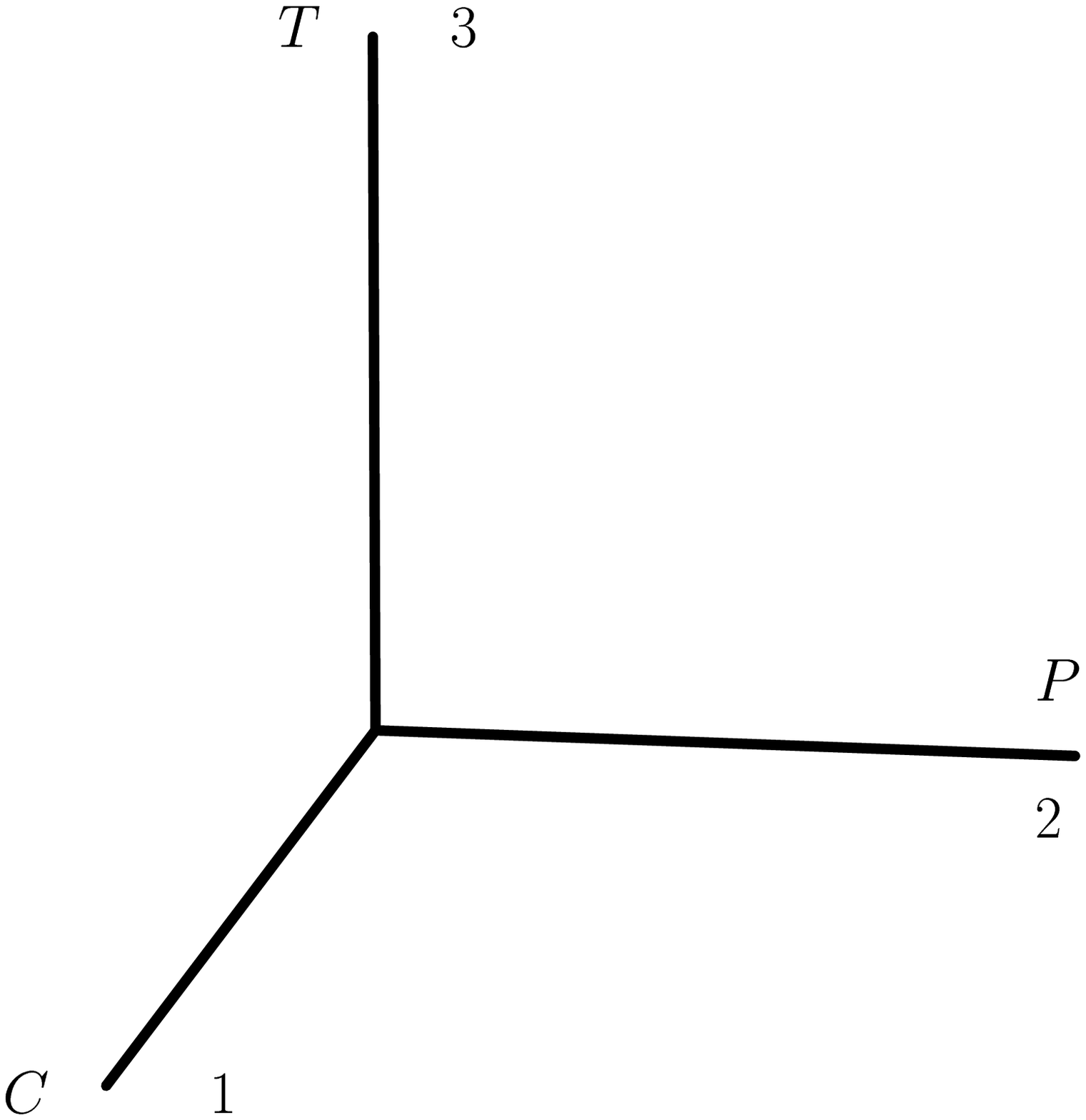}

%\bigskip

Fig.1

\bigskip

\includegraphics[width=50mm]{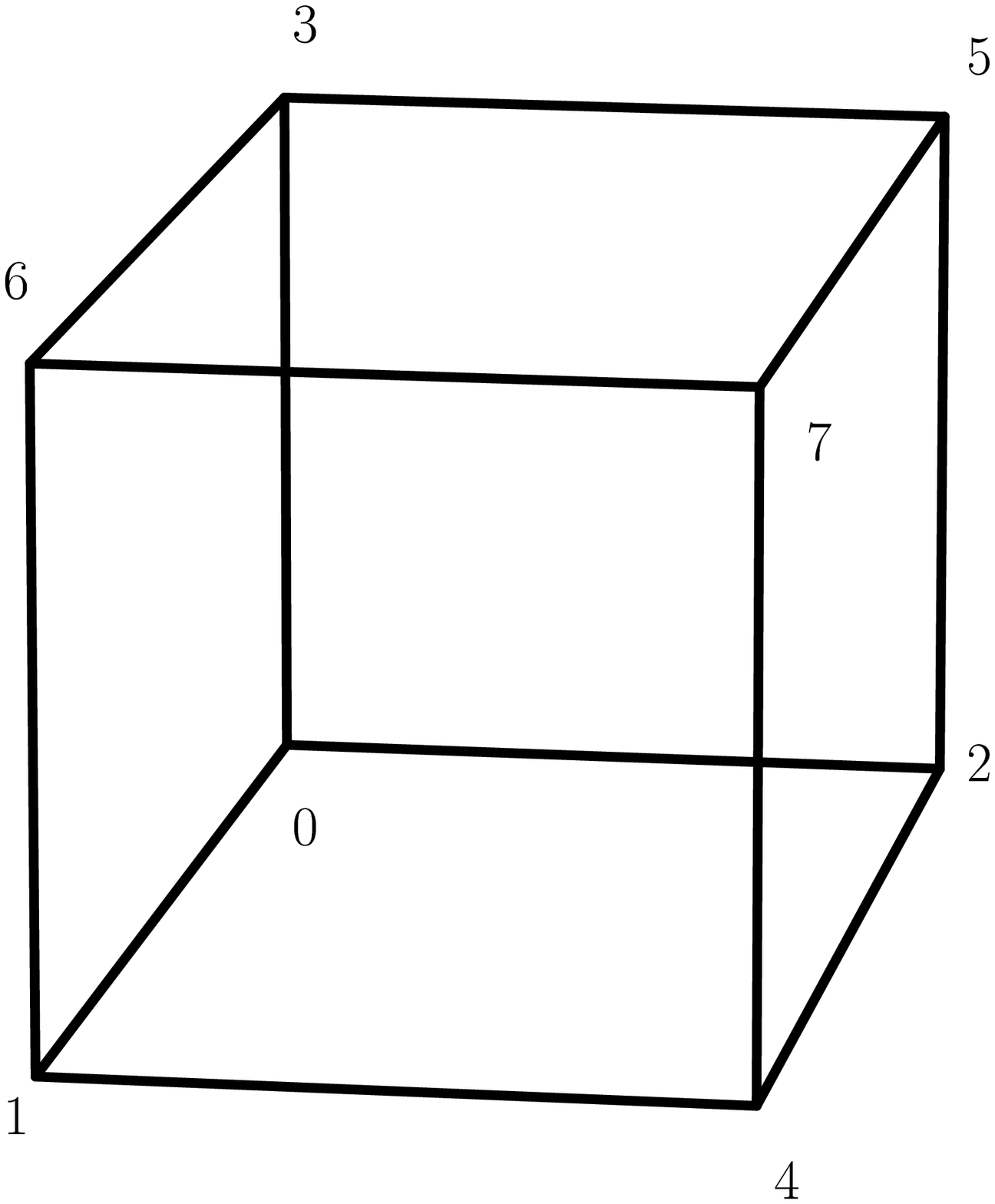}

%\bigskip

Fig.2

\bigskip

\includegraphics[width=80mm]{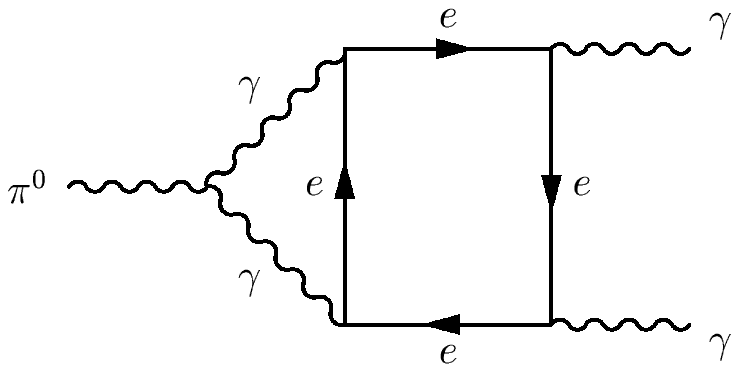}

\bigskip

Fig.3

\newpage

\bigskip

\includegraphics[width=80mm]{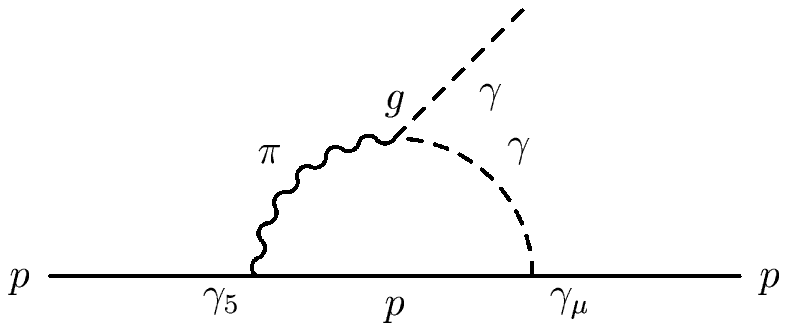}

\bigskip

Fig.4a

\bigskip

\includegraphics[width=80mm]{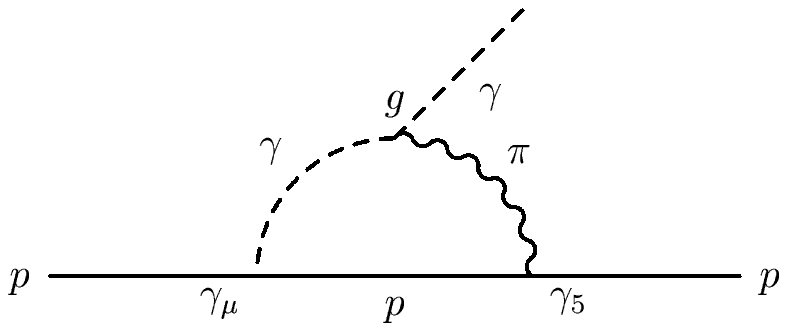}

\bigskip

Fig.4b

\bigskip

\includegraphics[width=80mm]{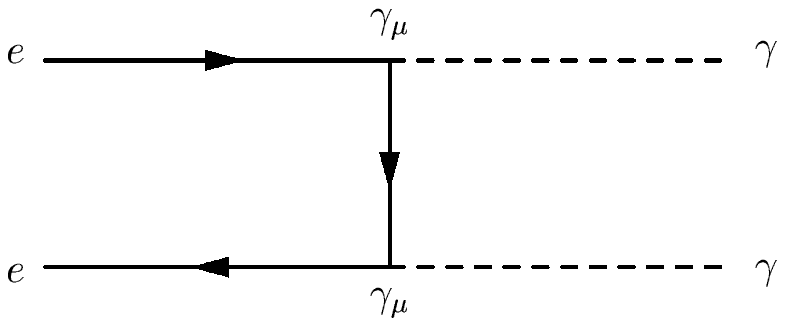}

\bigskip

Fig.5a

\bigskip

\includegraphics[width=80mm]{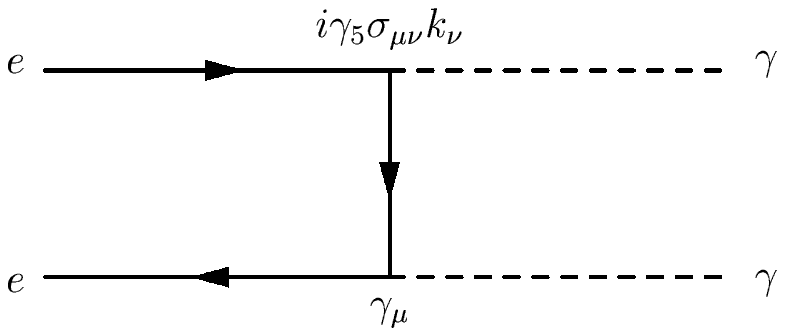}

\bigskip

Fig.5b

\bigskip

\includegraphics[width=80mm]{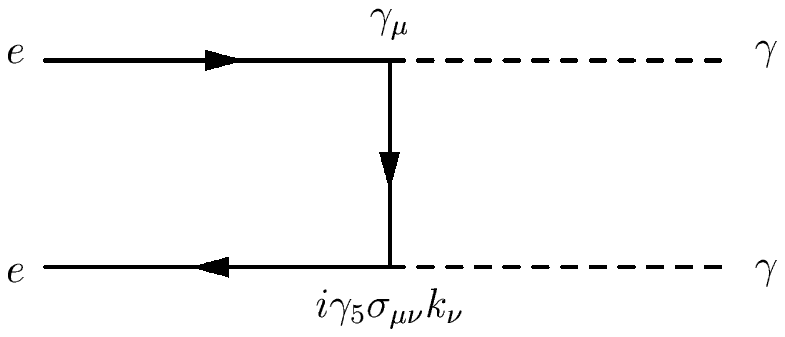}

\bigskip

Fig.5c

\end{center}

\end{document}